\documentclass[pra,twocolumn,showpacs,floatfix,a4paper]{revtex4}
\usepackage{bm,color,graphicx,amsmath,txfonts,amssymb}

\newcommand{\rp}[1]{(\ref{#1})}

\newcommand{\abs}[1]{\left|{#1}\right|}

\newcommand{\av}[1]{\left\langle #1 \right\rangle}

\newcommand{\al}[1]{^{(#1)}}
\newcommand{\da}{^\dagger}

\newcommand{\pt}[1]{\left( #1 \right)}
\newcommand{\pq}[1]{\left[ #1 \right]}
\newcommand{\pg}[1]{\left\{ #1 \right\}}

\newcommand{\bs}[1]{\boldsymbol #1}

\newcommand{\lpq}[1]{\left[ #1 \right.}

\newcommand{\rpq}[1]{\left. #1 \right]}

\newcommand{\ee}{{\rm e}}
\newcommand{\ii}{{\rm i}}

\newcommand{\nn}{{\nonumber}}

\newcommand{\mat}[2]{
                      \begin{array}{#1}
                       #2
                       \end{array}  }

\newcommand{\pp}[2]{ {\mbox{\scriptsize$
                      \begin{array}{c}
                       #1\\
                       #2
                      \end{array}$} }    }

\newcommand{\vbe}{{\bs \beta}}

\newcommand{\CC}{{\cal C}}

\newcommand{\GG}{{\cal G}}

\begin{document}

\title{Generation and detection of large and robust entanglement \\
between two different mechanical resonators in cavity optomechanics}

\author{J. Li, I. Moaddel Haghighi, N. Malossi, S. Zippilli, and D.~Vitali}
\affiliation{School of Science and Technology, Physics Division, University of Camerino, via Madonna delle Carceri, 9, I-62032 Camerino (MC), Italy, and INFN, Sezione di Perugia, Italy}

\begin{abstract}
We investigate a general scheme for generating, either dynamically or in the steady state, continuous variable entanglement between two mechanical resonators with different frequencies. We employ an optomechanical system in which a single optical cavity mode driven by a suitably chosen two-tone field is coupled to the two resonators. Significantly large mechanical entanglement can be achieved, which is extremely robust with respect to temperature.
\end{abstract}


\date{\today}
\maketitle

\section{Introduction}

Entanglement is the distinguishing feature of quantum mechanics and is the physical phenomenon according to which only the properties
of the entire system have precise values, while the physical properties of a subsystem can be assigned only in reference to those
of the other ones. It is now intensively studied because it corresponds to peculiar nonlocal correlations which
allows performing communication and computation tasks with an
efficiency which is not achievable classically \cite{Nielsen}.

Furthermore, for a deeper understanding of the boundary between the classical and quantum world,
it is important to investigate up to which macroscopic scale one can observe quantum behavior, and in particular under which conditions entanglement
between macroscopic objects, each containing a large number of the
constituents, can arise. Entanglement between two atomic ensembles
has been successfully demonstrated in Ref.~\cite{juuls01}, while entanglement between two Josephson-junction qubits has been detected in Refs.~\cite{Berkley,Steffen}. More recently, macroscopic entanglement has been demonstrated in electro-mechanical systems~\cite{Palomaki}: continuous variable (CV) entanglement, similar to that considered by Einstein Podolski and Rosen (EPR)~\cite{epr}, has been generated and detected between the position and momentum of a vibrational mode of a $15$ $\mu$m-diameter Al membrane, and the quadratures of a microwave cavity field, following the theory proposal of Ref.~\cite{hofer}.

Entanglement between two mechanical resonators (MRs) has been instead demonstrated only at the microscopic level, in the case of two trapped ions~\cite{Jost}, and between two single-phonon excitations in nano-diamonds~\cite{walmsley}. The realization of this kind of entanglement at the more macroscopic level of micromechanical resonators would be extremely important both for practical and fundamental reasons. In fact, on the one hand, entangled MRs at distant sites could represent an important building block for the implementation of quantum networks for long-distance routing of quantum information~\cite{Weedbrook}; on the other hand, these nonclassical states represent an ideal playground for investigating and comparing decoherence theories and modifications of quantum mechanics at the macroscopic level~\cite{marsh,romero-isart,ulbricht}.

Many different schemes have been proposed in the literature for entangling two MRs, especially exploiting optomechanical and electromechanical devices~\cite{amo,rmp}, in which the two MRs simultaneously interact with one or more electromagnetic cavity fields. Refs.~\cite{PRL02,Peng03,epl} considered the steady state of
different systems of driven cavities: Ref.~\cite{PRL02} focused on
two mirrors of a ring cavity, while Ref.~\cite{Peng03} assumed to drive two
independent linear cavities with two-mode squeezed light transferring its
entanglement to the cavity end-mirrors. Ref.~\cite{epl} instead considered a
double-cavity scheme in which one cavity couples to the relative motion of two MRs, and the second cavity
to their center-of-mass; when the system
is appropriately driven by squeezed light, such squeezing is transferred to the two MRs which are then prepared in a stationary EPR-like state. Actually, steady-state entanglement can be achieved, even if at a smaller value, also without squeezed driving, either between two movable mirrors in a Fabry-Perot cavity~\cite{jopa}, between two mechanical modes of a single movable mirror~\cite{genesnjp}, or in the case of two semi-transparent membranes interacting with two driven cavity modes~\cite{hartmann}.

A different approach for generating entangled MRs exploits conditional measurements on light modes entangled or correlated with mechanical degrees of freedom~\cite{entswap,bjorke,mehdi1,woolley,mehdi2,Savona}. In this case, entanglement is generated at the measurement and it has a finite lifetime which may be severely limited by the interaction of the MRs with their reservoirs. A similar strategy has been provided to enhance the entanglement of two MRs~\cite{JieNJP}. More recent proposal applied reservoir engineering ideas~\cite{poyatos,davidovich,zoller,cirac,pielawa} to optomechanical scenarios, by exploiting suitable multi-frequency drivings and optical architectures in order to achieve more robust generation of steady state entanglement between two MRs~\cite{Clerk,tan1,Tan,schmidt,WoolleyClerk,Abdi2,Buchmann}, eventually profiting from mechanical nonlinearities and/or parametric driving~\cite{bowen-clerk,nori}.

In the present paper we propose a novel optomechanical/electromechanical scheme for the generation of remarkably large CV entanglement between two MRs with different frequencies, which is also extremely robust with respect to thermal noise. The scheme is particularly simple, involving only a single, bichromatically-driven, optical cavity mode, and optimally works in a rotating wave approximation (RWA) regime where counter-rotating, non-resonant, terms associated with the bichromatic driving are negligible. The scheme shares some analogies with the reservoir-engineering schemes of Refs.~\cite{Clerk,Tan,WoolleyClerk,Buchmann}, but it may be used to generate robust entanglement also in a pulsed regime, in the special case of equal effective couplings at the two sidebands, where the system becomes analogous to the S{\o}rensen-M{\o}lmer scheme for entangling trapped ions in a thermal environment~\cite{SM}. This latter scheme has been already considered in an optomechanical scenario by Kuzyk {\it et al.}~\cite{Kuzyk} for entangling dynamically two optical modes via their common interaction with a single MR.

The paper is organized as follows. In Section II we derive the effective quantum Langevin equations (QLE) describing the dynamics of the system in the RWA.
In Section III we solve the dynamics in terms of the mechanical Bogoliubov modes of the system~\cite{Clerk,Tian,Tan}, derive the steady state of the system in the stable case, and provide simple analytical expressions for the achievable mechanical entanglement, showing its remarkable robustness with respect to temperature. In Section IV we instead consider the special case of equal couplings, when the system can be mapped to the S{\o}rensen-M{\o}lmer scheme~\cite{SM}, in which mechanical entanglement is generated only dynamically and slowly decays to zero at long times. In Section V we solve and discuss the exact dynamics of the system in order to establish the conditions under which the RWA does not seriously affect the robust generation of large mechanical entanglement. In Section VI we discuss the experimental detection of such entanglement and present some concluding remarks. In the Appendices we provide some detail on the dynamical evolution of the system, and present a careful derivation of the linearized QLE in the RWA regime.

\begin{figure*}[t]
\centering
\includegraphics[width=5.7in]{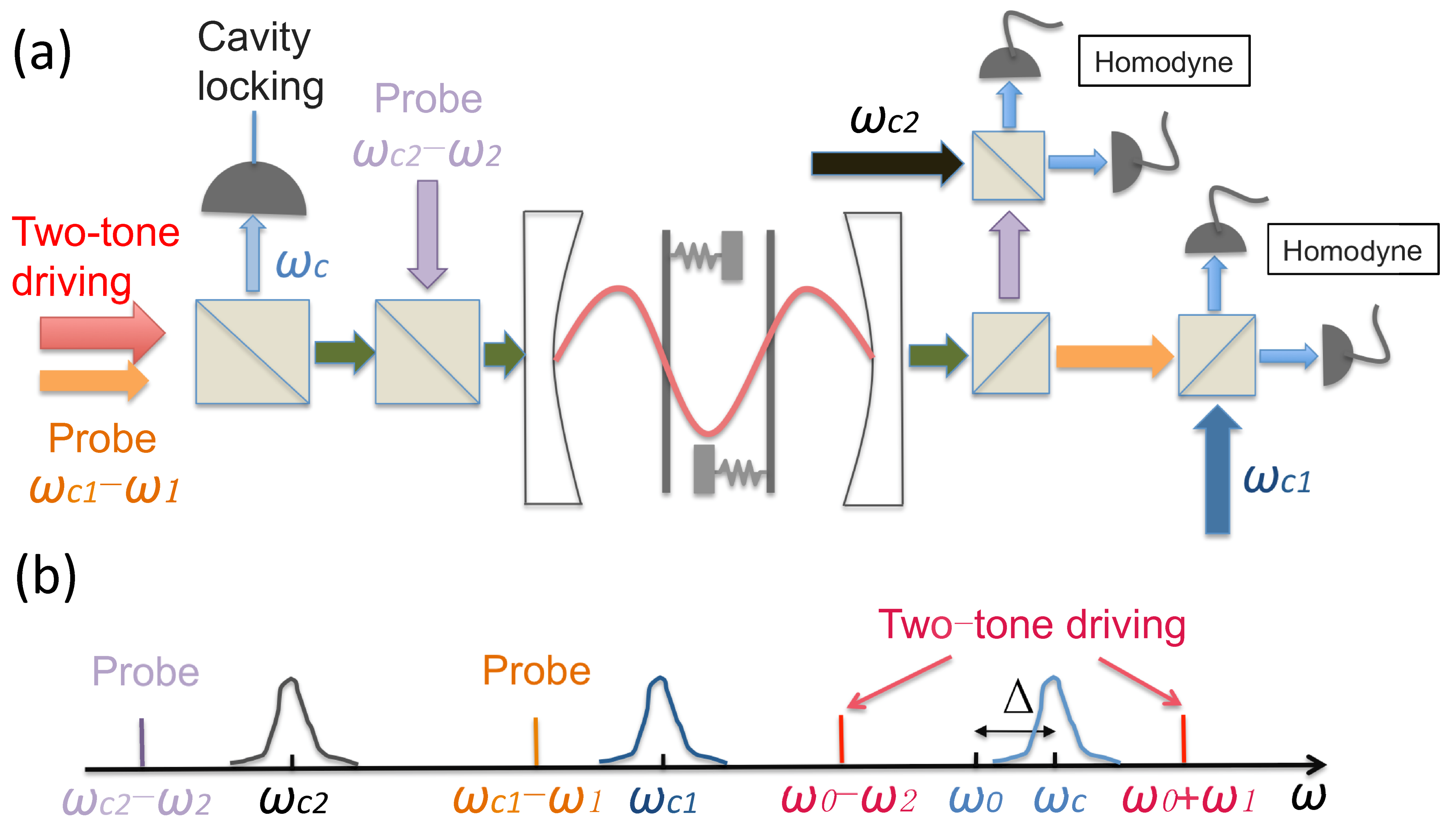}
\caption{Sketch of the proposed entanglement generation and detection scheme {\bf (a)}, and of the various pump and probe laser frequencies {\bf (b)}. The cavity mode is bichromatically driven at the two frequencies $\omega_0+\omega_1$ and $\omega_0-\omega_2$. Large and robust entanglement of the two mechanical resonators can be generated either dynamically or in the steady state. Two weak probe fields with detuning $\Delta^p_{j}=\omega_{cj}-\omega^p_{j}=\omega_{j}$, $j=1,2$, are then sent into the cavity. By homodyning the probe mode outputs, the mechanical quadratures $(x_j,p_j)$ are therefore measured, which allows one to construct the correlation matrix of the quadratures from which entanglement can be derived in a straightforward way.}
\label{detect}
\end{figure*}

\section{System Hamiltonian and derivation of the effective Langevin equations}\label{system}

As shown in Fig.~\ref{detect}, we consider an optical cavity mode with resonance frequency $\omega_c$ and annihilation operator $\hat{a}$ interacting via the usual optomechanical interaction with two different MRs, with frequencies $\omega_1$ and $\omega_2$ and annihilation operators $\hat{b}_1$ and $\hat{b}_2$ respectively. The cavity mode is bichromatically driven at the two frequencies $\omega_0+\omega_1$ and $\omega_0-\omega_2$, with the reference frequency $\omega_0$ detuned from the cavity resonance by a quantity $\Delta_0=\omega_c-\omega_0$. If we describe the cavity field in a reference frame rotating at the frequency $\omega_0$, then the system Hamiltonian is given by
\begin{eqnarray}
&&\hat{H}=\hbar\omega_1 \hat{b}^\dagger_1 \hat{b}_1+\hbar\omega_2 \hat{b}^\dagger_2 \hat{b}_2+ \hbar\Delta_0  \hat{a}^\dagger \hat{a} \nonumber \\
&& + \hbar\left[g_1\left(\hat{b}_1+\hat b_1^\dagger\right)+g_2\left(\hat{b}_2+\hat b_2^\dagger\right)\right]\hat{a}^\dagger \hat{a} \nonumber \\
&&+\hbar\pq{\pt{E_1 e^{-i\omega_1t}+E_2 e^{i\omega_2t}}\hat{a}^\dagger +{\rm H.C.}}\ .
\label{haml}
\end{eqnarray}
This means that the cavity mode is simultaneously driven on the blue sideband associated with the MR with annihilation operator $\hat{b}_1$, and on the red sideband associated with the MR with $\hat{b}_2$. The nonzero detuning $\Delta_0$ makes the present scheme different from the one studied in the supplementary material of Ref.~\cite{Clerk} which restricts to the resonant case $\Delta_0=0$. Our model is instead related to the scheme proposed by Kuzyk {\it et al.}~\cite{Kuzyk} for entangling dynamically two optical modes via their common interaction with a single MR: here we will dynamically entangle two MRs via their common interaction with an optical mode.

The system dynamics can be efficiently studied by linearizing the optomechanical interaction in the limit of large driving field.
In this case the average fields for both cavity, $\alpha(t)$, and mechanical degrees of freedom, $\beta_j(t)$, are large, and one can simplify the interaction Hamiltonian at lowest order in the field fluctuations
\begin{eqnarray}\label{displacement}
\delta \hat a(t)&=&\hat a(t)-\alpha(t),
\nn\\
\delta \hat b_j(t)&=&\hat b_j(t)-\beta_j(t)\ .
\end{eqnarray}
Differently from the typical optomechanical settings in which the steady state average fields are time-independent, here the bichromatic driving induces a time-dependent, periodic steady state average field which, in turn, implies time-dependent effective coupling strengths for the linearized dynamics of the fluctuations.
As originally discussed in~\cite{Mari}, and detailed in Appendix~\ref{linearization}, approximated dynamical equations for the fluctuation operators $\delta \hat a(t)$ and  $\delta \hat b_j(t)$ can be derived, in the interaction picture with respect to the Hamiltonian
$\hat{H}_0=\hbar\pt{\omega_1 \hat{b}^\dagger_1 \hat{b}_1+\omega_2 \hat{b}^\dagger_2 \hat{b}_2}$, by neglecting the non-resonant/time-dependent components of the effective linearized interactions.
It is possible to prove that this approach is justified when (see Eq.~\rp{cond1})
\begin{eqnarray}\label{cond01}
\abs{g_j\,\frac{E_j}{\omega_j}},\,\kappa \ll\omega_j,\abs{\omega_1-\omega_2}\ .
\end{eqnarray}
The corresponding QLE
including thermal noise and dissipation at rates $\kappa$ and $\gamma_j$ for the cavity and the mechanical mode $j\in1,2$ respectively, are
\begin{eqnarray}
 && \delta \dot{\hat{a}}=-\left(\kappa + i \Delta\right)\delta \hat{a}-i G_1 \delta \hat{b}_1^\dagger -i G_2 \delta \hat{b}_2+\sqrt{2\kappa} \hat{a}^{\rm in}\label{deltaa2}\\
 &&\delta\dot{\hat{b}}_1 =-\frac{\gamma_1}{2} \delta \hat{b}_1  -i G_1 \delta \hat{a}^{\dagger}+\sqrt{\gamma_1} \hat{b}_1^{\rm in},\\
&&\delta\dot{\hat{b}}_2 =-\frac{\gamma_2}{2} \delta \hat{b}_2  -i G_2^* \delta \hat{a}+\sqrt{\gamma_2} \hat{b}_2^{\rm in}, \label{b2st}
\end{eqnarray}
where
\begin{eqnarray}\label{G1G2}
G_1&=&\frac{g_1\,E_1}{\omega_1-\Delta+i\kappa},\nn\\
G_2&=&-\frac{g_2\, E_2}{\omega_2+\Delta-i\kappa},
\end{eqnarray}
are the (generally complex) linear optomechanical couplings, and $\hat{a}^{\rm in}$ and $\hat{b}_j^{\rm in}$ are standard input noise operators with zero mean, whose only nonzero correlation functions are $\av{\hat{a}^{\rm in}(t)\, \hat{a}^{\rm in}(t')\da}=\delta(t-t')$, $\av{\hat{b}_j^{\rm in}(t)\, \hat{b}_j^{\rm in}(t')\da}=(\bar n_j+1)\delta(t-t')$ and $\av{\hat{b}_j^{\rm in}(t)\da\, \hat{b}_j^{\rm in}(t')}=\bar n_j\delta(t-t')$, where $\bar{n}_j=\left[\exp\left(\hbar \omega_j/k_B T\right)-1\right]^{-1}$ is the mean thermal phonon number of the $j$-th MR, which we assume to stay at the same environmental temperature $T$.
Moreover, we note that here the new cavity detuning $\Delta$ includes the time-independent frequency shift induced by the optomechanical interaction, proportional to the DC component of the average mechanical oscillation amplitude $\beta_j(t)$, that we here denote with $\beta_j^{\rm DC}$ (see Appendix~\ref{linearization}). Specifically
 \begin{eqnarray}\label{barDelta}
\Delta=\Delta_0+2\sum_{j=1,2} g_j{\rm Re}\pq{\beta_j^{\rm DC}}\ .
\end{eqnarray}

We will see that the dynamics described by these equations allows to generate large and robust entanglement between the two MRs, either in the steady state or, in a particular parameter regime, during the time evolution with a flat-top pulse driving. We first notice that the system is stable when all the eigenvalues associated with the linearized dynamics of Eqs.~(\ref{deltaa2})-(\ref{b2st}) have negative real parts. The stability condition is quite involved in the general case, but it assumes a particularly simple form in the case of equal mechanical dampings, $\gamma_1=\gamma_2=\gamma$. In such a case, the system is stable if and only if
\begin{equation}\label{stab}
    |G_2|^2 > |G_1|^2-\frac{\kappa \gamma}{2}\left[1+\frac{4\Delta^2}{\left(\gamma+2\kappa\right)^2}\right].
\end{equation}
This stability condition reduces to the one derived in the supplementary material of Ref.~\cite{Clerk} in the case $\Delta=0$. We see that a nonzero detuning generally helps in keeping the system stable.

\section{Dark and bright Bogoliubov modes}
The coherent dynamics corresponding to the Eqs.~\rp{deltaa2}--\rp{b2st}, is described
by the effective linearized Hamiltonian
\begin{eqnarray}
&&\hat{H}_{\rm eff}=\hbar \Delta  \delta \hat{a}^\dagger \delta \hat{a}  +\hbar\left(G_1\delta \hat b_1^\dagger + G_2\delta \hat{b}_2\right)\delta\hat{a}^\dagger \nonumber \\
&&+ \hbar\left(G_1^*\delta \hat b_1+ G_2^*\delta \hat{b}_2^\dagger \right)\delta\hat{a}. \label{heff}
\end{eqnarray}
We can always adjust the phase reference of each MR (which will be determined by a local oscillator which must be used to measure the mechanical quadratures for verifying entanglement) so that we can take both $G_1$ and $G_2$ real.

Eq.~(\ref{heff}) naturally suggests to introduce two effective mechanical modes allowing to simplify the system dynamics. We assume for the moment $G_2>G_1$, which is a sufficient condition for stability (see Eq.~(\ref{stab})), and define
\begin{eqnarray}
\hat{\beta}_1&=&  \frac{G_2\delta \hat{b}_1 + G_1\delta \hat{b}_2^\dagger}{{\cal G}}=\delta \hat{b}_1 \cosh r  + \delta \hat{b}_2^\dagger \sinh r , \label{bet1}\\
\hat{\beta}_2&=&  \frac{G_2\delta \hat{b}_2 + G_1\delta \hat{b}_1^\dagger}{{\cal G}}=\delta \hat{b}_2 \cosh r  + \delta \hat{b}_1^\dagger \sinh r , \label{bet2}
\end{eqnarray}
where
\begin{equation}\label{calg-r}
    {\cal G}=\sqrt{G_2^2-G_1^2}, \;\;\;\; \tanh r = \frac{G_1}{G_2}.
\end{equation}
 Eqs.~(\ref{bet1})-(\ref{bet2}) define a Bogoliubov unitary transformation of the mechanical mode operators, which can also be written as
\begin{eqnarray}\label{twomode1}
\hat{\beta}_{1,2}&=&e^{-r \left(\delta \hat{b}_1^\dagger \delta \hat{b}_2^\dagger-\delta \hat{b}_1 \delta \hat{b}_2\right)}\delta \hat{b}_{1,2}e^{r \left(\delta \hat{b}_1^\dagger \delta \hat{b}_2^\dagger-\delta \hat{b}_1 \delta \hat{b}_2\right)}
\nn\\ &=&
\hat{S}(r)\delta \hat{b}_{1,2}\hat{S}(-r),
\end{eqnarray}
with $\hat{S}(r)$ the two-mode squeezing operator.
The Bogoliubov mode $\hat{\beta}_1$ describes the ``mechanical dark mode'', which does not appear in $H_{\rm eff}$, i.e., is decoupled from the cavity mode and therefore is a constant of motion in the absence of damping, while $\hat{\beta}_2$ is the ``bright'' mode interacting with the cavity mode.
This is equivalent to say that the dark mode $\hat{\beta}_1$ is the normal mode of the Hamiltonian dynamics with eigenvalue equal to zero. The other two normal modes of the system will be linear combinations of $\hat{\beta}_2$ and $\delta \hat{a}$. The Bogoliubov mode description has been already employed in cavity optomechanics, associated to two optical modes in Refs.~\cite{Clerk,Tian}, and to two mechanical modes in Refs.~\cite{Tan,WoolleyClerk} (see Appendix A for a derivation of the normal modes of the system and a study of its Hamiltonian dynamics).

\subsection{Stationary entanglement for different couplings}\label{Sec:stst}

For a realistic description of the system dynamics we must include cavity decay and mechanical dissipation.
It is convenient to rewrite the QLE in terms of the Bogoliubov modes, which in the case when $\gamma_1=\gamma_2\equiv \gamma$ assume the simple form
\begin{eqnarray}
 && \delta \dot{\hat{a}}=-\left(\kappa + i \Delta\right)\delta \hat{a}-i {\cal G} \hat{\beta}_2+\sqrt{2\kappa} \hat{a}^{\rm in}\label{deltaa3}\\
 &&\dot{\hat{\beta}}_1 =-\frac{\gamma}{2} \hat{\beta}_1  +\sqrt{\gamma} \hat{\beta}_1^{\rm in},\\
&&\dot{\hat{\beta}}_2 =-\frac{\gamma}{2} \hat{\beta}_2  -i {\cal G} \delta \hat{a}+\sqrt{\gamma} \hat{\beta}_2^{\rm in},\label{beta2}
\end{eqnarray}
where $\hat{\beta}_j^{\rm in}$, $j=1,2$, are two correlated thermal noise operators
whose only nonzero correlation functions are
\begin{eqnarray}
\av{\hat{\beta}_j^{\rm in}(t)\,\hat{\beta}_j^{\rm in}(t')\da}&=& \left[ \bar{n}^{\rm eff}_j(r)+1\right]\delta(t-t'),
\\
\av{\hat{\beta}_j^{\rm in}(t)\da\,\hat{\beta}_j^{\rm in}(t')}&=& \bar{n}^{\rm eff}_j(r)\delta(t-t'),
\nn\\
\av{\hat{\beta}_1^{\rm in}(t)\,\hat{\beta}_2^{\rm in}(t')}&=&\av{\hat{\beta}_1^{\rm in}(t)\da\,\hat{\beta}_2^{\rm in}(t')\da}=\bar m(r)\delta(t-t'),\nn
\end{eqnarray}
with the effective mean thermal phonon numbers
\begin{eqnarray}\label{neff1}
    \bar{n}^{\rm eff}_1(r) &=& \bar{n}_1 \cosh^2 r + (\bar{n}_2+1)\sinh^2 r , \\
    \bar{n}^{\rm eff}_2(r) &=& \bar{n}_2 \cosh^2 r + (\bar{n}_1+1)\sinh^2 r ,  \label{neff2}
\end{eqnarray}
and the inter-mode correlation
\begin{eqnarray}\label{barm}
\bar m(r)=\cosh r\,\sinh r \pt{\bar n_1+\bar n_2+1}\ .
\end{eqnarray}
If $\gamma_1 \neq \gamma_2$ a dissipative coupling term between the two Bogoliubov modes appears, which however does not have relevant effects because it is proportional to $|\gamma_1-\gamma_2|$ which is typically very small with respect to all other damping rates.

The dynamics associated with Eqs.~(\ref{deltaa3})--(\ref{beta2}) is simple: the bright mechanical mode $\hat{\beta}_2$ is cooled by the cavity, while the correlated reservoir create finite correlations between dark and bright modes.
In particular, the matrix of correlation for the vector of operators $\vbe=\pt{\hat{\beta}_1,\hat{\beta}_2,{\hat{\beta}_1}{}\da,{\hat{\beta}_2}{}\da}$, whose elements are $\pg{\CC_\beta}_{j,k}=\av{\pg{\vbe}_j\ \pg{\vbe}_k}$ is given, at the steady state, by
\begin{eqnarray}\label{CCbeta}
\CC_\beta=\pt{\mat{cccc}{
0 & \bar m_\beta & \bar n_1^{\rm eff} +1 & 0 \\
 \bar m_\beta  & 0 & 0 &\bar n_2^{\rm cool} +1  \\
 \bar n_1^{\rm eff}  & 0 & 0 & \bar m_\beta^*  \\
 0 & \bar n_2^{\rm cool}  & \bar m_\beta^* & 0
}} \ ,
\end{eqnarray}
with the number of excitation of the cooled bright mode and the correlations between the two Bogoliubov modes respectively given by
\begin{eqnarray}\label{occu2}
\bar{n}^{\rm cool}_2(r) &=&n_2^{\rm eff}(r)\pq{1-\frac{\pt{1-\epsilon}\, C_-}{1+\delta^2+C_-}},
\end{eqnarray}
and
\begin{eqnarray}\label{mbeta}
\bar m_\beta(r) =\bar m(r)\frac{2(1+i\delta)}{2(1+i\delta)+\pt{1-\epsilon}\, C_-}\ ,
\end{eqnarray}
where
\begin{eqnarray}\label{Cm}
C_-&=&\frac{2\GG ^2}{\gamma\kappa},\\
\epsilon&=&\frac{\gamma}{\gamma+2\kappa},\\
\delta&=&\frac{2\Delta}{\gamma+2\kappa}\ ,
\end{eqnarray}
and $C_-$ can be seen as an effective collective optomechanical cooperativity. The steady state correlation matrix can be expressed in terms of the original modes $b_1$ and $b_2$ by inverting the Bogoliubov transformation introduced in Eqs.~\rp{bet1} and \rp{bet2}. The result is\begin{eqnarray}
\CC_b&=&U\ \CC_\beta\ U^T \nn\\
&=&\pt{\mat{cccc}{
0 & \bar m_b& \bar n_{b1}+1 & 0 \\
\bar m_b & 0 & 0 & \bar n_{b2}+1  \\
 \bar n_{b1}  & 0 & 0 & \bar m_b \\
 0 & \bar n_{b2}  & \bar m_b & 0
}} \ ,
\end{eqnarray}
with
\begin{eqnarray}
U=\pt{\mat{cccc}{
\cosh r& 0 & 0 &-\sinh r \\
0& \cosh r& -\sinh r&0\\
0& -\sinh r& \cosh r&0\\
-\sinh r& 0& 0&\cosh r
 }} \ ,
\end{eqnarray}
and where now
\begin{eqnarray}
\bar n_{b1}&=&\bar n_1^{\rm eff}+\sinh^2 r \pt{1 + \bar n_1^{\rm eff}  + \bar n_2^{\rm cool} }
\nn\\
&&- 2\cosh r\ \sinh r\ {\rm Re} \pt{ \bar m_\beta},
\nn\\
\bar n_{b2}&=&\bar n_2^{\rm cool}+\sinh^2 r \pt{1 + \bar n_1^{\rm eff}  + \bar n_2^{\rm cool} }
\nn\\&&
- 2\cosh r\ \sinh r\ {\rm Re} \pt{ \bar m_\beta},
\nn\\
\bar m_{b}&=&\cosh^2 r\ \bar m_\beta+\sinh^2 r\ \bar m_\beta^* \nn\\
&& - \cosh r\ \sinh r \pt{1 + \bar n_1^{\rm eff}  + \bar n_2^{\rm cool} } \ .
\end{eqnarray}
The entanglement between modes $b_1$ and $b_2$, measured by means of the logarithmic negativity~\cite{eisert,plenio}, can be easily expressed in terms of these matrix elements as~\cite{Zippilli15}
\begin{eqnarray}\label{ENexact}
E_N&=&{\rm max}\pq{0,-\ln\, \nu},
\nn\\
\nu&=&1+\bar n_{b1}+\bar n_{b2}-\sqrt{4\abs{\bar m_b}^2+\pt{\bar n_{b1}-\bar n_{b2}}^2}\ .
\end{eqnarray}
When the collective cooperativity $C_-$ is sufficiently large, i.e., $C_- \gg \bar m(r)$, then $\bar m_\beta(r)$ is negligible (see Eq.~\rp{mbeta}).
This is the working regime in which we are particularly interested, because in this case, the second Bogoliubov mode can be cooled close to its ground state ($\bar{n}^{\rm cool}_2(r)\ll n_2^{\rm eff}(r)$), corresponding to an entangled state for the original mechanical modes. In this case the steady state correlation matrix for the Bogoliubov modes, in Eq.~\rp{CCbeta}, reduces to the correlation matrix of a
state given by the product of two thermal states with occupancies $\bar{n}^{\rm eff}_1(r)$ and $\bar{n}^{\rm cool}_2(r)$ respectively. For the two MR of interest, associated with the operator $\hat{b}_1$ and $\hat{b}_2$, such a state is just a two-mode squeezed thermal state~\cite{marian}
\begin{equation}\label{twomode2}
    \hat{\rho}_{1,2}=\hat{S}(r)\hat{\rho}_{\bar{n}^{\rm eff}_1(r),{\rm th}}\otimes \hat{\rho}_{\bar{n}^{\rm cool}_2(r),{\rm th}}\hat{S}(-r),
\end{equation}
where $\hat{S}(r)$ is given in Eq.~(\ref{twomode1}), and
\begin{equation}\label{therm}
   \hat{\rho}_{\bar{n},{\rm th}}=\sum_{n=0}^{\infty}\frac{\bar{n}^n}{(1+\bar{n})^{n+1}}|n\rangle \langle n|
\end{equation}
is the density matrix of the thermal equilibrium state of a resonator with occupancy $\bar{n}$. Such a state is entangled for sufficiently large $r$ and not too large mean thermal excitation number.

\begin{figure}[b]
\centering
\includegraphics[width=2.8in]{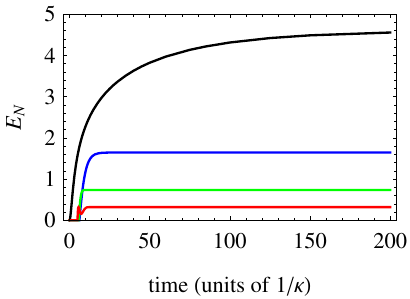}
\caption{Time evolution of the logarithmic negativity $E_N$ starting from an initial uncorrelated state with the optical mode fluctuations $\delta \hat{a}$ in the vacuum state and each MR in its thermal state with mean phonon number: i) $\bar{n}_1=\bar{n}_2=0$, $G_1=0.995 G_2$ (black line); ii) $\bar{n}_1=200$, $\bar{n}_2=100$, $G_1=0.918 G_2$ (blue line); iii) $\bar{n}_1=1000$, $\bar{n}_2=500$, $G_1=0.82 G_2$ (green line); $\bar{n}_1=2000$, $\bar{n}_2=1000$, $G_1=0.75 G_2$ (red line); the other parameters are $\gamma=10$ s$^{-1}$, $\kappa=10^5$ s$^{-1}$, $G_2=10^5$ s$^{-1}$, $\Delta = 10^3$ s$^{-1}$.}%
\label{entdyn1}
\end{figure}

This prediction of large stationary entanglement is confirmed in Fig.~\ref{entdyn1}, where we plot the time evolution of the entanglement between the two MRs, quantified in terms of the logarithmic negativity $E_N$, obtained from the solution of Eqs.~(\ref{deltaa2})-(\ref{b2st}). Figure 2 refers to an experimentally achievable set of parameters, $\gamma=10$ s$^{-1}$, $\kappa=10^5$ s$^{-1}$, $G_2=10^5$ s$^{-1}$, $\Delta = 10^3$ s$^{-1}$, and to different values of mean thermal phonon numbers $\bar{n}_1$, $\bar{n}_2$, and of the ratio $G_1/G_2$. We see that remarkable values of $E_N$ are achieved at low temperatures, and that stationary mechanical entanglement is quite robust with respect to temperature because one has an appreciable value of $E_N \simeq 0.32$ even for $\bar{n}_1=2000$, $\bar{n}_2=1000$. The time to reach the steady state is essentially given by the inverse of the cooling rate of the bright Bogoliubov mode, which is approximately given by $t_s\simeq(\kappa^2+\Delta^2)/(\GG^2\kappa)$ (see Eqs.~(\ref{deltaa3})--(\ref{beta2})).

Eq.~(\ref{twomode2}) suggests that one could achieve large stationary entanglement between the two MRs by taking a large two-mode squeezing parameter $r$, and a large collective cooperativity $C_-\gg1$ in order to significantly cool the bright Bogoliubov mode. However the corresponding optimization of the system parameters, and especially of the two couplings $G_1$ and $G_2$, is far from being trivial. In fact, $r$ increases when $G_1 \to G_2$, which however implies, at a fixed value of $G_2$, a decreasing value of ${\cal G}$ and therefore of $C_-$
(see Eq.~(\ref{calg-r}) and Eq.~(\ref{Cm})); moreover increasing $r$ has also the unwanted effect of increasing $\bar m_\beta(r)$
that is the correlations between the two Bogoliubov modes (see Eqs.~\rp{barm} and \rp{mbeta}).

\begin{figure}[th]
\centering
\includegraphics[width=2.8in]{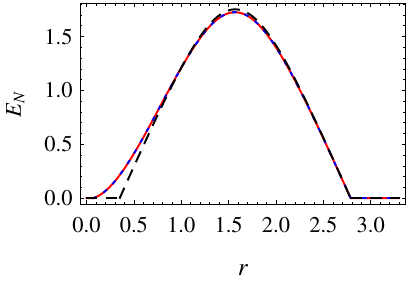}
\caption{$E_N$ at the steady state versus $r$ for $\gamma=10$ s$^{-1}$, $\kappa=10^5$ s$^{-1}$, $G_1=10^5$  s$^{-1}$, $\Delta=0$, implying a cooperativity $C_1 =2 \times 10^4$, and $\bar{n}_1=200$, $\bar{n}_2=100$. The full red line refers to the steady state solution of the QLE in Eqs.~\rp{deltaa3}--\rp{beta2}, that is given by
Eq.~\rp{ENexact}, the blue dashed line
is evaluated with the approximated value of $\nu$ reported in Eq.~\rp{nubeta0},
and the black dashed line corresponds to the approximation in Eq.~\rp{nuappr}.
}
\label{logneg1}
\end{figure}

However, a judicious choice of parameters is possible, allowing to get very large stationary mechanical entanglement, even in the presence of non-negligible values of the thermal occupancies $\bar{n}_1$ and $\bar{n}_2$. At a given value of $G_1$, this is obtained by taking a sufficiently large value of the associated single-mode cooperativity, $C_1=2G_1^2/\kappa \gamma \gg 1$, and correspondingly optimizing the value of $G_2$, i.e., of $r$. In fact, the logarithmic negativity associated with the stationary state of Eq.~(\ref{twomode2}) can be evaluated in terms of the parameter
\begin{eqnarray}\label{nubeta0}
\nu\Big|_{\bar m_\beta\to 0}&=&\pq{\bar n_+(r)+1}\pt{\cosh^2 r+\sinh^2 r}
\\&&-\sqrt{\bar n_-(r)^2+4\pq{\bar n_+(r)+1}^2\sinh^2 r\cosh^2 r}\ , \nn
\label{lognegtmsts1}
\end{eqnarray}
where $\bar{n}_{\pm}(r)=\bar{n}^{\rm eff}_1(r)\pm \bar{n}^{\rm cool}_2(r)$, and  $\bar{n}^{\rm cool}_2(r)$ can be explicitly rewritten in terms of the cooperativity $C_1$ as
\begin{eqnarray}\label{ncool2}
    \bar{n}^{\rm cool}_2(r)&=&\left[\bar{n}_2 \cosh^2 r + (\bar{n}_1+1)\sinh^2 r \right]
  \nn\\&&\times
    \pq{1-\frac{\pt{1-\epsilon}\, C_1}{\sinh^2 r\pt{1+\delta^2}+C_1}}\ .
\end{eqnarray}
The dependence of $E_N$ versus $r$, for given values of $C_1$, $\bar{n}_1$ and $\bar{n}_2$, shows a maximum and then decays to zero for large $r$ (see Fig.~\ref{logneg1} which refers to $C_1 =2\times10^4$ and $\bar{n}_1=200$, $\bar{n}_2=100$). This behavior is described by a very simple approximated expression
valid in the limit $C_1\gg\ee^{2 r}\gg\ee^{-2r}$, with not very large $\bar{n}_{1,2}$, and when $\delta,\epsilon\to 0$ (corresponding to $\gamma,\Delta\ll\kappa$),
\begin{eqnarray}\label{nuappr}
\nu\sim 2\ee^{-2r}+\pt{1+\bar n_1+\bar n_2}\frac{\ee^{2r}}{4C_1}
\end{eqnarray}
which exhibits a minimum (hence corresponding to maximum entanglement) as a function of $r$ at
\begin{equation}\label{rmax}
    r\simeq r^{\rm opt}=\frac{1}{4}\ln\left(\frac{8C_1}{\bar{n}_1+\bar{n}_2+1}\right)\ ,
\end{equation}
given by
$\nu^{\rm opt}\sim\sqrt{\frac{2(1+\bar n_1+\bar n_2)}{C_1}}\  .$
For values of $r$ much larger or much smaller than this value, the resonators may not be entangled.
When $r$ is increased to very large values $r\gg r^{\rm opt}$, $\GG$ is reduced and the cooling dynamics becomes slow as compared to the standard mechanical dissipation, which takes place at rate $\sim\gamma\ (\bar n_{j}+1)$, so that the correlations between the MRs cannot be efficiently generated. On the other hand, at small $r\ll r^{\rm opt}$ the Bogoliubov modes are essentially equal to the original modes, so that the cavity cools only the second resonator, and also in this case mechanical entanglement can not  be observed.
Fig.~\ref{logneg1} also shows that the simplified expression of Eq.~(\ref{nuappr}) provides a simple but valid approximation for large $C_1$ and a very good estimate of the optimal value of the two-mode squeezing parameter $r$, i.e., of $G_1/G_2$, given by Eq.~\rp{rmax}.
The corresponding value of the logarithmic negativity is
\begin{eqnarray}\label{ENapprox}
E_N\sim\frac{1}{2}\ln\left[\frac{C_1}{2\pt{1+\bar n_1+\bar n_2}}\right]
\end{eqnarray}
and shows that once that the ratio $G_1/G_2$ is optimized, the achievable stationary entanglement between the two MRs increases with increasing $C_1/(\bar{n}_1+\bar{n}_2)$.

The above analysis of the stationary entanglement of the two MRs extends the results of Ref.~\cite{Clerk} in various directions. First of all, our model extends to the case of nonzero detuning $\Delta$ a model discussed in the Supplementary material of Ref.~\cite{Clerk}. We see that a nonzero detuning has a limited effect of the dynamic of entanglement generation, providing only an effective increase of $\bar n_2^{\rm cool}$, which however becomes negligible as soon as $\Delta \ll \kappa$ (see Eq.~(\ref{ncool2})). Moreover, Ref.~\cite{Clerk} provided an explicit expression for $E_N$ only for the case of negligible thermal occupancies and not too large values of $r$, while the present discussion applies for arbitrary values of $r$, $\bar{n}_1$ and $\bar{n}_2$.

\section{Dynamical evolution in the case of equal couplings}

In the special case of equal couplings $G_1=G_2 \equiv G$, i.e., ${\cal G}=0$, the Bogoliubov modes cannot be defined anymore and the description of the preceding Section cannot be applied. The dynamics is nonetheless interesting and still allows for the generation of appreciable entanglement between the two MRs, even though only at finite times and not in the stationary state. We notice that in this special case, our scheme becomes analogous to that of Ref.~\cite{Kuzyk}, that showed that two appropriately driven optical modes can be entangled with a pulsed scheme by their common interaction with a MR. More precisely, the QLE of Eqs.~(\ref{deltaa2})-(\ref{b2st}) are the same as those studied in Ref.~\cite{Kuzyk} but now referred to two mechanical modes coupled to the same optical mode, i.e., with exchanged roles between optical and mechanical degrees of freedom.

The physical mechanism at the basis of the generation of dynamical entanglement can be understood by looking at the Hamiltonian evolution of the system at equal couplings. Such mechanism essentially coincides with the one proposed for entangling internal states of trapped ions by Milburn~\cite{milburn} and by S{\o}rensen and M{\o}lmer~\cite{SM}, and first applied to an optomechanical setup by Kuzyk {\it et al.}~\cite{Kuzyk}. In the present case, the common interaction with the bichromatically driven optical mode dynamically entangles the two MRs, and at special values of the interaction time the optical mode is decoupled from the two MRs and mechanical entanglement can be strong.

\begin{figure}[b]
\centering
\includegraphics[width=2.8in]{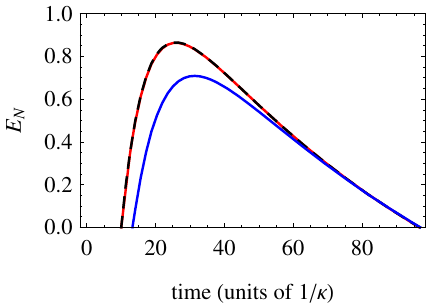}
\caption{Time evolution of $E_N$ in the case of equal couplings for the parameter set $\gamma=10$ s$^{-1}$, $\kappa=10^5$ s$^{-1}$,  $G=10^5$ s$^{-1}$, $\bar{n}_1=200$, $\bar{n}_2=100$, and for three different values of the detuning: $\Delta=10^3$ s$^{-1}$ (black dashed line), $\Delta=10^4$ s$^{-1}$ (red full line), and $\Delta=10^5$ s$^{-1}$ (blue full line).}%
\label{SMdynamics}%
\end{figure}

At equal couplings it is convenient to rewrite the effective Hamiltonian after linearization of Eq.~(\ref{heff}) in terms of mechanical and optical quadratures, using the expressions $\delta \hat{b}_j=(\hat{x}_j+i\hat{p}_j)/\sqrt{2}$, $j=1,2$, and $\delta\hat{a}=(\hat{X}+i\hat{Y})/\sqrt{2}$. One gets
\begin{equation}
\hat{H}_{\rm eff}=\frac{\hbar \Delta}{2}  \left(\hat{X}^2+\hat{Y}^2\right) + \hbar G \sqrt{2} \left(\hat{x}_+ \hat{X}-\hat{p}_-\hat{Y}\right),\label{heffsm}
\end{equation}
where $\hat{x}_{\pm}=(\hat{x}_1\pm\hat{x}_2)/\sqrt{2}$, $\hat{p}_{\pm}=(\hat{p}_1\pm\hat{p}_2)/\sqrt{2}$ are linear combinations of the two position and momentum operators of the two MRs. The Heisenberg evolution of these latter mechanical operators can be solved in a straightforward way, by exploiting the fact that $\hat{x}_+$ and $\hat{p}_-$ are two commuting conserved observables. One gets (see also Refs.~\cite{Kuzyk,milburn,SM})
\begin{eqnarray}
 &&\hat{x}_+(t)=\hat{x}_+(0),\;\;\;\; \hat{p}_-(t)=\hat{p}_-(0), \\
  && \hat{x}_-(t)= \hat{x}_-(0) +\frac{2G^2}{\Delta^2}\left(\sin \Delta t -\Delta t\right)\hat{p}_-(0) \nonumber \\
  &&+ \frac{2G^2}{\Delta^2}\left(1-\cos \Delta t\right)\hat{x}_+(0) \\
  && -\frac{G\sqrt{2}}{\Delta}\sin\Delta t \hat{Y}(0)+ \frac{G\sqrt{2}}{\Delta}\left(1-\cos \Delta t\right) \hat{X}(0),\nonumber \\
   && \hat{p}_+(t)= \hat{p}_+(0) -\frac{2G^2}{\Delta^2}\left(\sin \Delta t -\Delta t\right)\hat{x}_+(0) \nonumber \\
   &&+ \frac{2G^2}{\Delta^2}\left(1-\cos \Delta t\right)\hat{p}_-(0) \\
  && -\frac{G\sqrt{2}}{\Delta}\sin\Delta t \hat{X}(0)- \frac{G\sqrt{2}}{\Delta}\left(1-\cos \Delta t\right) \hat{Y}(0). \nonumber
\end{eqnarray}
Relevant interaction times are those when the MR dynamics decouple from that of the optical cavity, and this occurs at $t_m=2 m \pi/\Delta$, $m=1,2,\ldots$, where
\begin{eqnarray}
  && \hat{x}_-(t_m)= \hat{x}_-(0) -2m \pi \frac{2G^2}{\Delta^2}\hat{p}_-(0),  \\
   && \hat{p}_+(t)= \hat{p}_+(0) +2m \pi \frac{2G^2}{\Delta^2}\hat{x}_+(0).
\end{eqnarray}
This map describes a stroboscopic evolution in which the two MRs become more and more entangled, because it corresponds to the application of the unitary operator
\begin{eqnarray}\label{um}
   && U_m=\exp\left[-i\frac{2\pi G^2 m}{\Delta^2}\left(\hat{x}_+^2+\hat{p}_-^2\right)\right] \\
   &&= \exp\left[-i\frac{2\pi G^2 m}{\Delta^2}\left(\delta\hat{b}_1^{\dagger}\delta\hat{b}_1+\delta\hat{b}_2^{\dagger}\delta\hat{b}_2+1+\delta\hat{b}_1^{\dagger}\delta\hat{b}_2^{\dagger}+\delta\hat{b}_1\delta\hat{b}_2\right)\right] \nonumber .
\end{eqnarray}
This ideal behavior is significantly modified by the inclusion of damping and noise, especially the one associated with the cavity mode, which acts on the faster timescale $1/\kappa$ and seriously affects the cavity-mediated interaction between the two MRs, as soon as $\kappa$ becomes comparable to $\Delta$. Mechanical entanglement is large for large $G/\Delta$ and we expect well distinct peaks for $E_N$ at interaction times $t_m$, in the ideal parameter regime $G\gg \Delta \gg \kappa$. In the more realistic regime in which $G$, $\Delta$ and $\kappa$ are comparable, the peaks will be washed out, but we still expect an appreciable value for the mechanical entanglement for a large interval of interaction times. This is confirmed by the numerical solution of the time evolution associated with the QLE shown in Fig.~\ref{SMdynamics}, which refers to the parameter set $\gamma=10$ s$^{-1}$, $\kappa=10^5$ s$^{-1}$,  $G=10^5$ s$^{-1}$, $\bar{n}_1=200$, $\bar{n}_2=100$, and to three different values of the detuning, $\Delta=10^3$ s$^{-1}$ (black dashed line), $\Delta=10^4$ s$^{-1}$ (red full line), and $\Delta=10^5$ s$^{-1}$ (blue full line). We see that an appreciable value of $E_N$ (even though smaller than the one achievable at the same $\bar{n}_1$ and $\bar{n}_2$ after the optimization of $G_1/G_2$ of the previous Section) is reached for a large interval of interaction times $t$. Therefore even at equal couplings (and nonzero detuning) one can entangle the two resonators with a pulsed experiment. Mechanical entanglement instead vanishes in the stationary state.

\begin{figure*}[ht!]
\centering
\includegraphics[width=5.9cm]{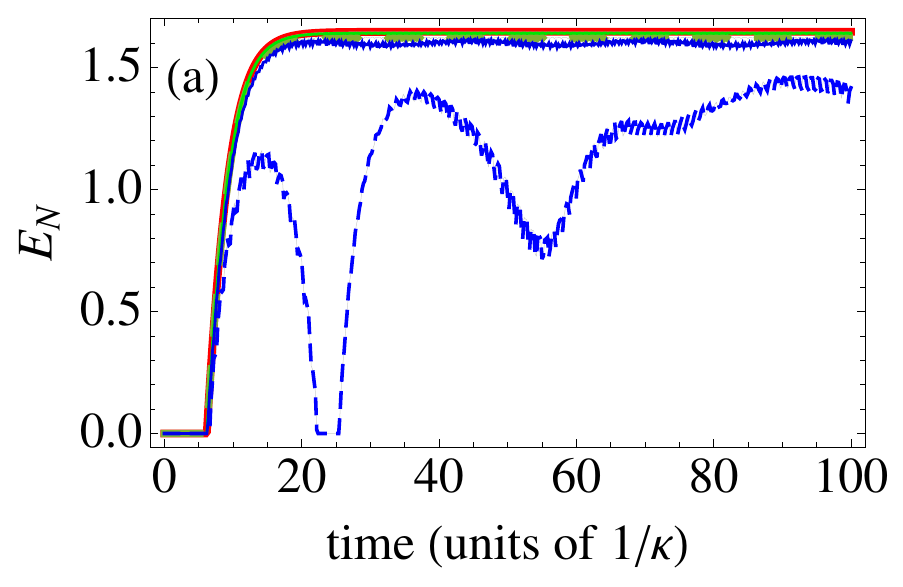}
\includegraphics[width=5.9cm]{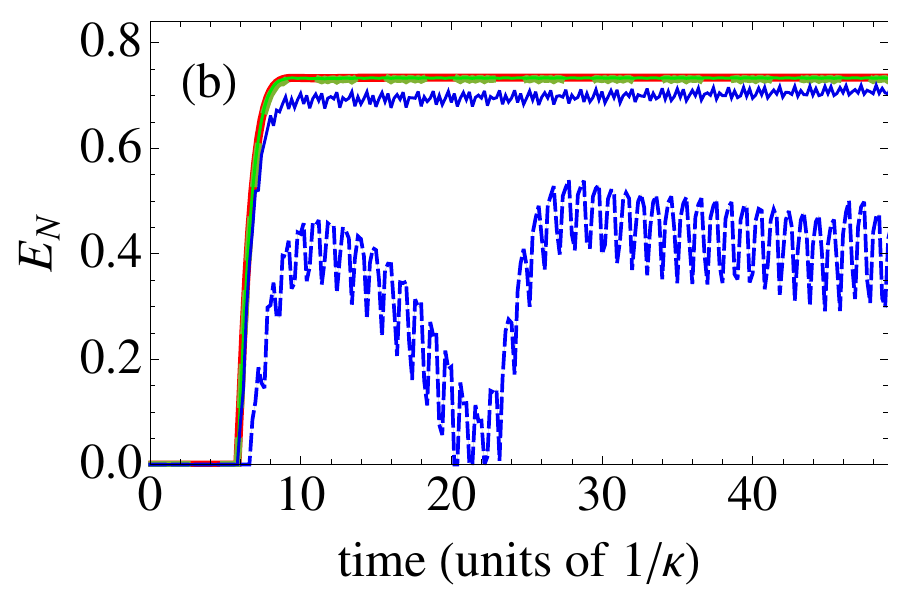}
\includegraphics[width=5.9cm]{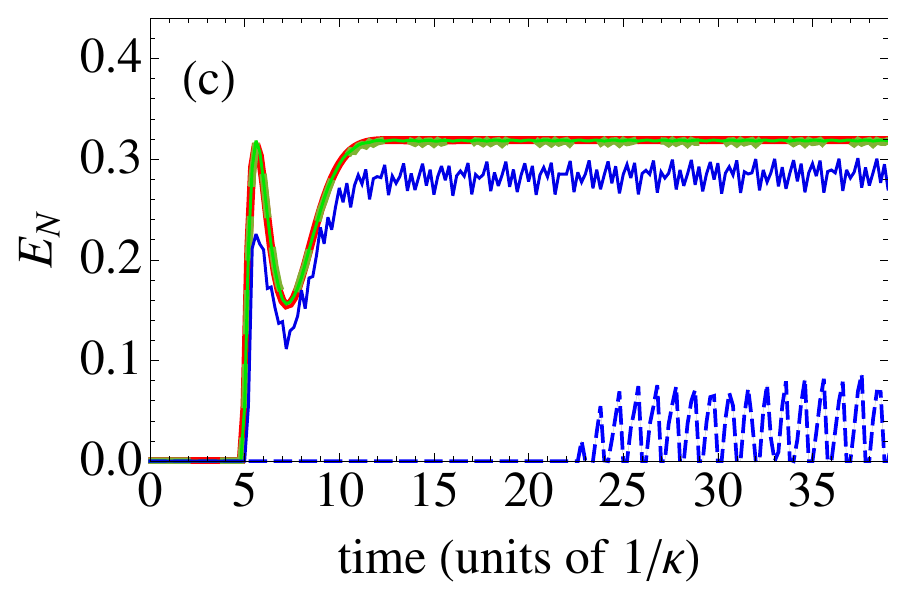}
\caption{Comparison of the time evolution of $E_N$ evaluated with and without the non-resonant terms, when $G_2>G_1$.
The red lines are evaluated with the model described by Eqs.~\rp{deltaa2}-\rp{b2st} that does not take into account the non-resonant terms. The green lines are evaluated with Eq.~\rp{qle1_b2}, which takes into account the non-resonant terms by considering the expansion of the steady state solutions of $\alpha(t)$ and $\beta_j(t)$, at first order in $g$ as defined in Eq.~\rp{firstorder}. The blue lines are evaluated instead with Eq.~\rp{qle1_b1} by considering the expansion for $\alpha(t)$ and $\beta_j(t)$, calculated iteratively with Eqs.~\rp{alphabeta_g}, \rp{chi}-\rp{solalphabeta}, up to sixth order in $g$; in particular these results take into account the full dynamics of the average fields $\alpha(t)$ and $\beta_j(t)$, with initial condition $\alpha(0)=\beta_j(0)=0$, and not only the steady state as in the case of the green lines.
The solid lines refer to $\omega_2=100\kappa$ and $\omega_1=50\kappa$, while the dashed lines refer to $\omega_2=50\kappa$ and $\omega_1=25\kappa$.  The other parameters are $G_2=\kappa$, $\Delta=0.01\kappa$, $\gamma=10^{-4}\kappa$, and $g=10^{-4}\kappa$.
Moreover in (a)
 $G_1=0.918\kappa$ $\bar n_1=200$, $\bar n_2=100$;
in (b)
 $G_1=0.82\kappa$, $\bar n_1=1000$, $\bar n_2=500$;
in (c)
 $G_1=0.75\kappa$, $\bar n_1=2000$, $\bar n_2=1000$.
}
\label{res1}
\end{figure*}
\begin{figure*}[ht!]
\centering
\includegraphics[width=5.9cm]{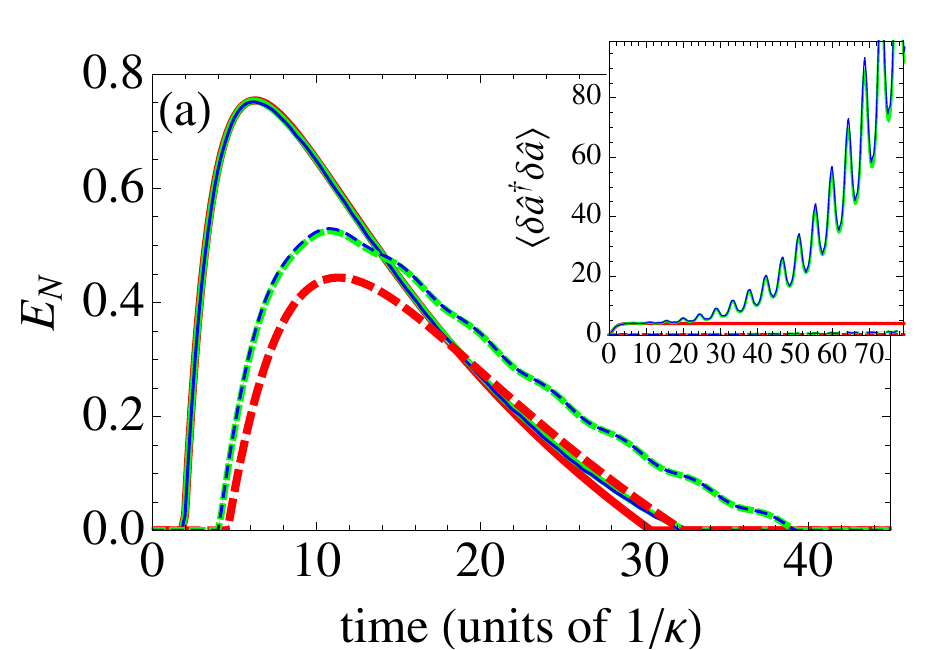}
\includegraphics[width=5.9cm]{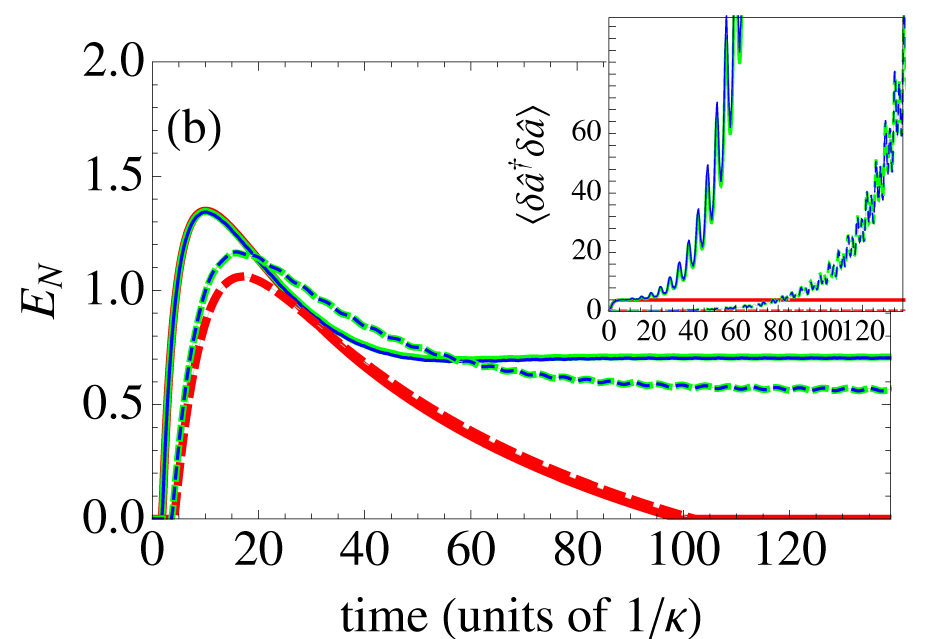}
\includegraphics[width=5.9cm]{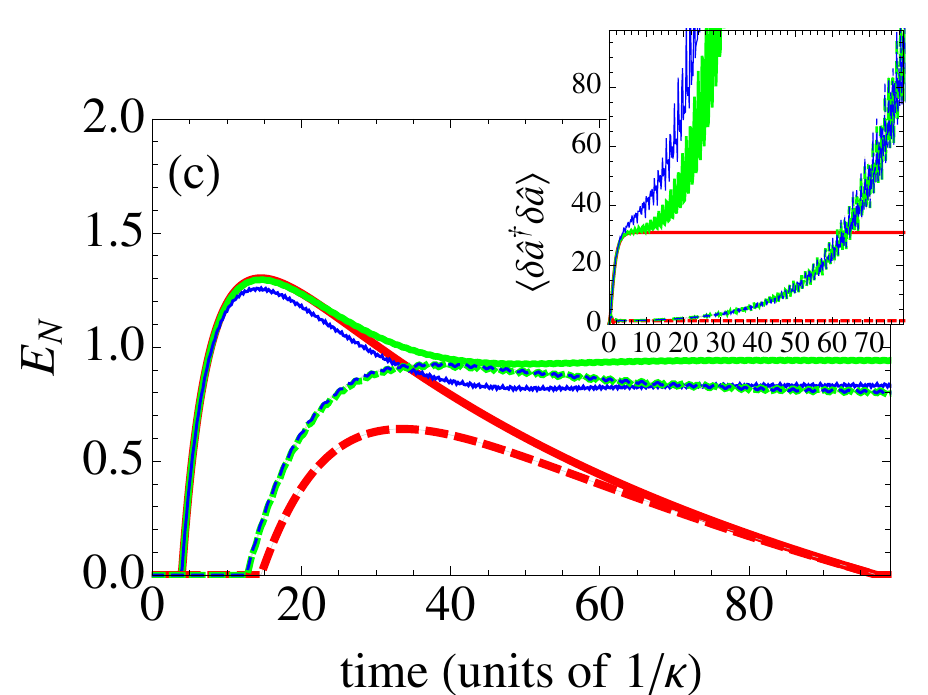}
\caption{Comparison of the time evolution of $E_N$ evaluated with and without the non-resonant terms, when $G_1=G_2$. The insets report the corresponding average photon number $\av{\delta\hat a\da\ \delta\hat a}$.
Red, green and blue lines are evaluated as in Fig.~\ref{res1}.
The solid lines are found for $\Delta=0.01\kappa$ and the dashed lines for $\Delta=5\kappa$. In (a) $\gamma=0.03\kappa$, $\omega_1=58\kappa$, $\omega_2=100\kappa$, $\bar n_1=2$ and $\bar n_2=1$; in (b) $\gamma=0.01\kappa$, $\omega_1=58\kappa$, $\omega_2=100\kappa$, $\bar n_1=2$ and $\bar n_2=1$; in (c) $\gamma=0.001\kappa$, $\omega_1=51\kappa$, $\omega_2=100\kappa$, $\bar n_1=20$ and $\bar n_2=10$.
The other parameters are $G_1=G_2=\kappa$ and $g=10^{-4}\kappa$.
}
\label{res2}
\end{figure*}

\section{Effect of the counter-rotating terms. Study of the exact dynamics}

The derivation of the effective linearized dynamics of Appendix~\ref{linearization} suggests that the counter-rotating terms that we have neglected may play an important role when the mechanical frequencies are not too large with respect to the other parameters (see also the comments in the supplementary material of Ref.~\cite{Clerk}). It is therefore interesting to study their effect by comparing the above predictions, both in the case of $G_2 > G_1$ and in the case of equal couplings, to the solution of the exact QLE obtained without neglecting the various time-dependent terms.

In Appendix~\ref{linearization} we describe the derivation of the effective linearized equations that we have studied in the preceding sections and that is based on the elimination of fast rotating terms and on the expansion of the linearized coupling strength at lowest order in $g_j$.
Here we analyze the limit of validity of these approximations by solving numerically the system dynamics  with the inclusion of the non-resonant terms expanded at different orders in powers of $g_j$.
In Fig.~\ref{res1} and \ref{res2} the red lines are evaluated without the non-resonant terms (i.e., the treatment of the preceding Sections), while the green and the blue ones take into account the full dynamics. In particular the green lines are computed by expanding the average fields $\alpha(t)$ and $\beta_j(t)$ (that have been introduced in Eq.~\rp{displacement} and discussed in Appendix~\ref{linearization}), at the lowest relevant order in powers of $g$, while for the blue ones they have been expanded at sixth order in powers of $g$. Moreover, the green line results are found considering only the steady state solution for $\alpha(t)$ and $\beta_j(t)$, while the blue lines are computed taking into account their full dynamics (that includes also the transient regime before the steady state is reached) with initial condition $\alpha(0)=\beta_j(0)=0$.

In Fig.~\ref{res1} we compare the time evolution of the entanglement evaluated with and without the time-dependent terms when $G_2>G_1$. The parameters used in these plots are consistent with those used in Fig.~\ref{entdyn1}. Specifically the three red curves in Figs.~\ref{res1} (a), (b) and (c), that are barely visible because almost entirely covered by the green curves, are equal to the three lowest curves in Fig.~\ref{entdyn1}.
We observe that the green and the red lines are always very close, meaning that the linearized RWA treatment is a very good approximation of the full dynamics when $\alpha(t)$ and $\beta_j(t)$ can be expanded at lowest order in $g$. Nevertheless, we note that if the mechanical frequencies are not large enough and higher order terms are taken into account together with the full dynamics of $\alpha(t)$ and $\beta_j(t)$, then the results can be significantly different as described by the blue curves. Specifically, the solid-blue lines are evaluated for sufficiently large values of the mechanical frequencies so that the condition in Eq.~\rp{cond01} is well fulfilled, and the effective linearized RWA dynamics recovers with significant accuracy the one determined with the inclusion of the non-resonant terms. The dashed-blue lines are instead evaluated for smaller frequencies. In this case it is evident that the non-resonant terms have a significant role in the system dynamics and that the lowest order expansion of the coefficients $\alpha(t)$ and $\beta_j(t)$ does not provide an accurate description. We note that according to Eq.~\eqref{cond01}, in order to eliminate the fast rotating terms, the ratios $\omega_j/G_j$ have to be much larger than one.  Although the dashed-blue curves are evaluated for a ratio $\omega_1/G_1$ of roughly 25, which can be considered significantly large, we have found, indeed, that it is not enough for a faithful approximation of the system dynamics with the model discussed in the preceding sections. The conclusive analysis of these cases would, possibly, require a non-perturbative approach that is beyond the scope of the present work. A final remark is in order. We have verified that the discrepancy between the dashed-blue lines and the red ones is due to the combined effect of the higher order terms and of the transient initial dynamics of $\alpha(t)$ and $\beta_j(t)$. Specifically, when we consider either the lowest order terms and the transient dynamics, or the higher order terms and only the steady state of $\alpha(t)$ and $\beta_j(t)$, the corresponding results for the entanglement dynamics are very similar to the red lines.

In Fig.~\ref{res2} we study the case of equal couplings $G_1=G_2$. In this case solid and dashed lines differ in the values of the cavity detuning $\Delta$. In general larger $\Delta$ (dashed lines) corresponds to smaller entanglement, and the results evaluated by including the counter-rotating terms tends to exhibit larger entanglement than the corresponding ones obtained without the non-resonant terms.
The solid curves are found with smaller $\Delta$. In this case red, green and blue lines are very close when the mechanical dissipation is sufficiently large as in Fig.~\ref{res2} (a). Larger discrepancies are found when the mechanical dissipation is reduced as in Fig.~\ref{res2} (b) and (c), especially at relatively large time. We observe in fact that, while the red curves for the entanglement decay to zero at large time, the corresponding green and blue lines seem to approach a finite sizable value.
As shown by the insets, when this different behaviour is observed, the average photon number in the cavity tends to diverge. This is a signature of the fact that the full dynamics including counter-rotating terms is actually unstable, even though the RWA dynamics without these terms is stable (see Eq.~\rp{stab}). We have confirmed the unstable nature of the time-dependent dynamics by calculating the Floquet exponents of the dynamical equations of the system.  In fact, when $\alpha(t)$ and $\beta_j(t)$ are considered in their steady state, one has a system of linear differential equations with periodic, time-dependent coefficients (see Appendix~\ref{linearization}), and the Floquet theory can be applied in this case~\cite{Teschl}; we have verified that for the parameters of Fig.~\ref{res2} there is always at least one positive Floquet exponent, meaning that the system is unstable.
This implies that, in general, the corresponding results are well-grounded only for relatively short time until the populations are not exceedingly large. On the other hand, our results show that in a pulsed experiment with the parameters of Fig.~\ref{res2}, these instabilities do not constitute a serious hindrance to the creation of significant entanglement at finite times.

Therefore, when the mechanical frequencies are sufficiently large ($\omega_j \gtrsim 10^2 \kappa$) (and, limited only to the case of equal couplings, when also mechanical damping is not too small), the effective linearized RWA dynamics obtained by neglecting the counter-rotating terms approximates with very good accuracy the full system dynamics.

\section{Strategies for the experimental detection of mechanical entanglement}

We finally discuss how to detect the generated mechanical entanglement between the two MRs at different frequencies. The present entanglement describes EPR-like correlations between the quadratures of the two MRs and therefore we need to perform homodyne-like detection of these quadratures. In the linearized regime we are considering the state of the two MRs is a Gaussian CV state, which is fully characterized by the matrix of all second-order correlations between the mechanical quadratures. Therefore from the measurement of these correlations one can extract the logarithmic negativity $E_{N}$.
One does not typically have direct access to the mechanical quadratures, but one can exploit the currently available possibility to perform low-noise and highly efficient homodyne detection of optical and microwave fields, and implement an efficient transfer of the mechanical phase-space quadratures onto the optical/microwave field.

As suggested in Ref.~\cite{wien} and then implemented in the electromechanical entanglement experiment of Ref.~\cite{Palomaki}, the motional quadratures of a MR can be read by homodyning the output of an additional ``probe'' cavity mode. In particular, if the readout cavity mode is driven by a
much weaker laser so that its back-action on the mechanical mode can
be neglected, and resonant with the first red sideband of the mode, i.e., with a detuning $\Delta^p_j=\omega_j$, $j=1,2$, the probe mode adiabatically follows the MR dynamics, and the output of the readout cavity $a_j^{out}$ is given by (see Fig.~\ref{detect})~\cite{wien}
\begin{equation}
\label{output} a_j^{out}= i \frac{G^p_j }{\sqrt{\kappa}}  \delta b_j +
a_{j}^{in},\;\;\;j=1,2,
\end{equation}
with $G^p_j$ the very small optomechanical coupling with the probe mode. Therefore using a probe mode for each
MR, changing the phases of the corresponding local oscillator, and
measuring the correlations between the probe mode outputs, one
can then detect all the entries of the correlation matrix and
from them numerically extract the logarithmic negativity
$E_{N}$.

\subsection{Concluding remarks}

We have studied in detail a general scheme for the generation of large and robust CV entanglement between two MRs with different frequencies through their coupling with a single, bichromatically driven cavity mode. The scheme extends and generalizes in various directions similar schemes exploiting driven cavity modes~\cite{Clerk,Tan,WoolleyClerk,Kuzyk} for entangling two MRs or two cavity modes. The scheme is able to generate a remarkably large entanglement between two macroscopic oscillators in the stationary state, i.e., with virtually infinite lifetime, and it is quite robust because one can achieve appreciably large CV entanglement even with thermal occupancies of the order of $10^3$. The scheme is particularly efficient in the limit where counter-rotating terms due to the bichromatic driving of the cavity mode are negligible, and we have verified with a careful numerical analysis that this is well justified when the two mechanical frequencies are sufficiently large $\omega_j \gtrsim 10^2 \kappa$.

\section{Acknowledgments}
This work has been supported by the European Commission (ITN-Marie Curie project cQOM and FET-Open Project iQUOEMS), by MIUR (PRIN 2011).

\appendix

\section{Normal modes and Hamiltonian dynamics}

It is straightforward to see that the diagonal form of the interaction Hamiltonian of Eq.~(\ref{heff}) is
\begin{eqnarray}
&&\hat{H}_{\rm eff}=\hbar \lambda_0  \hat{\beta}_1^\dagger \hat{\beta}_1  +\hbar \lambda_1  \hat{\alpha}_1^\dagger \hat{\alpha}_1+\hbar \lambda_2  \hat{\alpha}_2^\dagger \hat{\alpha}_2 \label{heffdia},
\end{eqnarray}
where
\begin{eqnarray}
\hat{\alpha}_1&=&  \cos \theta \hat{\beta}_2 + \sin \theta \delta \hat{a}, \\
\hat{\alpha}_2&=&  \cos \theta \delta \hat{a}- \sin \theta \hat{\beta}_2,
\end{eqnarray}
define the other two normal modes together with the dark mode $\hat{\beta}_1$, introduced in Eq.~\rp{bet1}, with $\theta$ defined by the condition $\tan 2\theta = -2{\cal G}/\Delta$, while the eigenvalues are given by $\lambda_0=0$, $\lambda_1=\left(\Delta-\tilde{\Delta}\right)/2$, $\lambda_2=\left(\Delta+\tilde{\Delta}\right)/2$, with $\tilde{\Delta}=\sqrt{\Delta^2+4 {\cal G}^2}$.

The normal modes allows to understand the dynamics in the absence of optical and mechanical damping processes. In fact, from Eq.~(\ref{heffdia}) one can easily derive the Heisenberg evolution of the mechanical bosonic operators. By inverting Eqs.~(\ref{bet1})-(\ref{bet2}) one has, $\delta\hat{b}_1(t)=\cosh r \hat{\beta}_1(t) - \sinh r \hat{\beta}_2^\dagger (t)$, $\delta\hat{b}_2(t)=\cosh r \hat{\beta}_2(t) - \sinh r \hat{\beta}_1^\dagger (t)$ and using $\alpha_j(t)=e^{i\lambda t}\alpha_j(0)$, $j=1,2$, and $\beta_1(t)=\beta_1(0)$, one gets
\begin{eqnarray}
  &&\delta\hat{b}_1(t) = \cosh r \hat{\beta}_1(0) \label{b1} \\
  &&- \sinh r \exp\left[i\frac{\Delta t}{2}\right]\left[\cos\frac{\tilde{\Delta}t}{2}-i\cos 2\theta \sin\frac{\tilde{\Delta}t}{2}\right]\hat{\beta}_2^\dagger (0) \nonumber \\
  &&-i\sinh r \sin 2\theta \exp\left[i\frac{\Delta t}{2}\right] \sin\frac{\tilde{\Delta}t}{2} \delta\hat{a}(0)\nonumber \\
 && \delta\hat{b}_2(t) = -\sinh r \hat{\beta}_1^{\dagger}(0) \label{b2} \\
&& + \cosh r \exp\left[-i\frac{\Delta t}{2}\right]\left[\cos\frac{\tilde{\Delta}t}{2}+i\cos 2\theta \sin\frac{\tilde{\Delta}t}{2}\right]\hat{\beta}_2(0) \nonumber \\
 && +i\cosh r \sin 2\theta \exp\left[-i\frac{\Delta t}{2}\right] \sin\frac{\tilde{\Delta}t}{2} \delta\hat{a}(0). \nonumber
\end{eqnarray}
We now look for special time instants at which the two mechanical modes can be strongly entangled. A necessary condition for such dynamical entanglement is that at these times, the cavity mode must be decoupled from the mechanical modes and Eqs.~(\ref{b1})-(\ref{b2}) show that it occurs when $\sin\tilde{\Delta}t/2=0$, i.e., $t_p=2\pi p/\sqrt{\Delta^2+4 {\cal G}^2}$, $p=1,2,\ldots$. At these time instants one has
\begin{eqnarray}
  &&\delta\hat{b}_1(t_p) = \left[\cosh^2 r -e^{i\phi_p}\sinh^2 r\right]\delta\hat{b}_1(0) \label{b1tp} \\
  && + \sinh r \cosh r \left(1-e^{i\phi_p}\right)\delta\hat{b}_2^\dagger (0) ,\nonumber \\
  &&\delta\hat{b}_2^{\dagger}(t_p) = \left[e^{i\phi_p}\cosh^2 r -\sinh^2 r\right]\delta\hat{b}_2^\dagger (0) \label{b2tp} \\
  &&- \sinh r \cosh r \left(1-e^{i\phi_p}\right) \delta\hat{b}_1(0),\nonumber
\end{eqnarray}
where $\phi_p=\pi p\left(1+\Delta/\tilde{\Delta}\right)$. In particular, if $e^{i\phi_p}=-1$ one gets
\begin{eqnarray}
  \label{b1tp2} &&\delta\hat{b}_1(t_p) = \cosh 2 r \delta\hat{b}_1(0)  + \sinh 2r \delta\hat{b}_2^\dagger (0) , \\
\label{b2tp2}  &&\delta\hat{b}_2^{\dagger}(t_p) = -\cosh 2 r \delta\hat{b}_2^\dagger (0) - \sinh 2 r  \delta\hat{b}_1(0),
\end{eqnarray}
i.e., the state of the two MRs at time $t_p$ is the result of the application of the two-mode squeezing operator with squeezing parameter $2r$, $\hat{S}(2r)$ (see Eq.~(\ref{twomode1})) to their initial state.
In the usual case of an initial thermal state for the two MRs with mean thermal phonon numbers $\bar{n}_j$, the state at time $t_p$ is therefore a two-mode squeezed thermal state~\cite{marian} (see Eq.~(\ref{twomode2})), with logarithmic negativity~\cite{eisert,plenio}
\begin{eqnarray}\label{lognegtmsts}
   && E_N(t_p)=-\frac{1}{2}\ln \left[\bar{n}_-^2+(\bar{n}_+ +1)^2 \cosh 8r \right. \\
  && \left. - \sqrt{(\bar{n}_+ +1)^4 \sinh^2 8r + 4 \bar{n}_-^2 (\bar{n}_+ +1)^2 \cosh^2 4r}\right] ,\nonumber
\end{eqnarray}
where $\bar{n}_{\pm}=\bar{n}_1\pm\bar{n}_2$. For the relevant case of not too small values of the squeezing parameter $r$, $E_N$ can be well approximated with its value at equal mean thermal phonon number $\bar{n}_- =0$,
\begin{equation}\label{lognegtmsts2}
    E_N(t_p) \simeq 4r -\ln\left[\bar{n}_+ +1\right],
\end{equation}
showing that at this interaction time, the entanglement between the MR can be very large, even if starting from a relatively hot state, by properly tuning the ratio $G_2/G_1$, i.e., the intensity of the two tones. This large mechanical entanglement is achieved when the condition $e^{i\phi_p}=-1$ is also satisfied for a given integer $p$. This is obtained for any odd $p$ when $\Delta =0$, or by properly adjusting the value of ${\cal G}$ for a given $\Delta \neq 0$, i.e., if
\begin{equation}\label{optg}
    {\cal G}^2_p = \Delta^2 \frac{d(2p-d)}{4(p-d)^2}\;\;\;d\;{\rm odd}, \;0<d<2p,\;d\neq p.
\end{equation}
This dynamical scheme for the generation of continuous variable mechanical entanglement is similar to the Bogoliubov scheme proposed in Ref.~\cite{Tian} for entangling two optical cavity modes. It is extremely hard however to use it for entangling two mechanical modes as in the present case, because the cavity decay rate is comparable to ${\cal G}$ and $\Delta$ in typical situations, thereby strongly affecting the ideal Hamiltonian dynamics described here.

\section{Linearization of the optomechanical dynamics with two-frequency drives}\label{linearization}

The system dynamics is described by the following QLE
\begin{eqnarray}
\dot{\hat{a}}&=&-\pq{\kappa + i \pt{\Delta_0+\omega_-}}\hat{a}-i\left[E_1 e^{-i\omega_+ t}+E_2 e^{i\omega_+ t}\right] +\sqrt{2\kappa} \hat{a}^{\rm in}\nn \\
&&-i \left[g_1\left(\hat{b}_1 +\hat b_1^\dagger \right)+g_2\left(\hat{b}_2+\hat b_2^\dagger \right)\right]\hat{a}, \nonumber \\
\dot{\hat{b}}_j& =&-\pt{\frac{\gamma_j}{2}+i\omega_{j}} \hat{b}_j-i g_j  \hat{a}^{\dagger} \hat{a} +\sqrt{\gamma_j} \hat{b}_j^{\rm in} \ \;\;\;j=1,2,\nn\\
\end{eqnarray}
where, here, differently from the description used in Sec.~\ref{system}, we are representing the cavity field in a reference frame rotating at the frequency $\omega_0-(\omega_2-\omega_1)/2$, and we have introduced the frequencies
$$\omega_\pm=\frac{\omega_2\pm\omega_1}{2}\ .$$
The other parameters and operators are defined in the main text.

If we perform a time dependent displacement, for both cavity and mechanical degrees of freedom, of the form
\begin{eqnarray}\label{displacement2}
\hat a(t)&=&\delta \hat a(t)+\alpha(t),
\nn\\
\hat b_j(t)&=&\delta \hat b_j(t)+\beta_j(t),
\end{eqnarray}
the QLE reduce to the form
\begin{eqnarray}\label{qle1}
\dot{\delta\hat{a}}&=&-\pg{\kappa + i \pt{\Delta_0+\omega_-}+2i g_1{\rm Re}\pq{\beta_1(t)}+2i g_2{\rm Re}\pq{\beta_2(t)}} \delta \hat{a}
\nn\\&&
-iA(t) +\sqrt{2\kappa} \hat{a}^{\rm in}\nn \\
&&
-i \alpha(t)\pq{g_1\pt{\delta\hat b_1+\delta\hat b_1\da}+g_2\pt{\delta\hat b_2+\delta\hat b_2\da}}
\nn\\
&&-i \left[g_1\left(\delta \hat{b}_1 +\delta \hat b_1^\dagger \right)+g_2\left(\delta \hat{b}_2+\delta \hat b_2^\dagger \right)\right]\delta \hat{a},  \\
\dot{\delta \hat{b}}_j& =&-\pt{\frac{\gamma_j}{2}+i\omega_{j}} \delta \hat{b}_j-iB_j(t)+\sqrt{\gamma_j} \hat{b}_j^{\rm in}
\nn\\&&
-ig_j\pq{\alpha(t)\delta\hat a\da+\alpha(t)^*\delta\hat a}-i g_j  \delta \hat{a}^{\dagger} \delta \hat{a},
\label{qle2}
\end{eqnarray}
where the new driving terms, $A(t)$ and $B_j(t)$ read
\begin{eqnarray}\label{AB}
A(t)&=&-i\left[E_1 e^{-i\omega_+ t}+E_2 e^{i\omega_+ t}\right]  -\frac{\partial \alpha(t)}{\partial t}
-\pq{\kappa + i\pt{ \Delta_0+\omega_-}}\alpha(t)
\nn\\&&
-2i\pt{ g_1{\rm Re}\pq{\beta_1(t)}+g_2{\rm Re}\pq{\beta_2(t)}} \alpha(t),
\nn\\
B_j(t)&=& -\frac{\partial \beta_j(t)}{\partial t}-\pt{\frac{\gamma_j}{2}+i\omega_{j}}\beta_j(t)-ig_j\abs{\alpha(t)}^2\ .
\end{eqnarray}
When $g_1$ and $g_2$ are sufficiently small and we chose $\alpha(t)$ and $\beta_j(t)$ such that $A(t)=0$
and $B_j(t)=0$, then the non-linear terms, i.e. the last terms in the two equations~\rp{qle1} and \rp{qle2}, can be neglected.  The equations $A(t)=0$ and $B_j(t)=0$ define a set of non-linear differential equations with periodic driving for the parameters $\alpha(t)$ and $\beta_j(t)$. The solution can be evaluated perturbatively in the small parameters $g_1$ and $g_2$~\cite{Mari}. Here we assume $g_1=g_2\equiv g$ and we observe that the solutions for $\alpha(t)$ and $\beta_j(t)$, with initial condition  $\alpha(0)=\beta_j(0)=0$, contain, respectively, only even and odd powers of $g$,
\begin{eqnarray}\label{alphabeta_g}
\alpha(t)&=&\sum_\pp{p=0}{p\ {\rm even}}^{\infty} g^{p}\alpha\al{p}(t),
\nn\\
\beta_j(t)&=&\sum_\pp{p=1}{p\ {\rm odd}}^{\infty} g^{p}\beta_j\al{p}(t)\ .
\end{eqnarray}
The equations for each component of these expansions can be written in the form
\begin{eqnarray}\label{dotalphabeta}
\dot\alpha\al{p}(t)&=&-z\,\alpha\al{p}(t)
+\Xi_\alpha\al{p}(t),
\nn\\
\dot\beta_j\al{p}(t)&=&-w_j\,\beta_j\al{p}(t)
+\Xi_\beta\al{p}(t),
\end{eqnarray}
where
\begin{eqnarray}\label{zw}
z&=&\kappa + i\pt{ \Delta_0+\omega_-},\nn\\
w_j&=&\frac{\gamma_j}{2}+i\omega_{j},
\end{eqnarray}
and the driving terms
are defined recursively as
\begin{eqnarray}\label{chi}
\Xi_\alpha\al{p}(t)
&=&-2 i \sum_{q=0}^{p-1}\alpha\al{q}(t)
\nn\\&&\times
\pt{{\rm Re}\pq{\beta_1\al{p-q-1}(t)}+{\rm Re}\pq{\beta_2\al{p-q-1}(t)}},
\nn\\
\Xi_\beta\al{p}(t)
&=&-i\sum_{q=0}^{p-1} \alpha\al{q}(t)\ {\alpha\al{p-q-1}(t)}^*\ ,
\end{eqnarray}
with the initial condition
\begin{eqnarray}
\Xi_\alpha\al{0}(t)&=&E_1\,\ee^{-\ii\,\omega_+\,t}+E_2\,\ee^{\ii\,\omega_+\,t},
\nn\\
\Xi_\beta\al{0}(t)&=&0\ .
\end{eqnarray}
In particular they can always be rewritten as sums of exponential functions of the form
\begin{eqnarray}
\Xi_\alpha\al{p}(t)&=&\sum_n\chi_\alpha\al{p,n}\,\ee^{\zeta_\alpha\al{p,n}\,t},
\nn\\
\Xi_\beta\al{p}(t)&=&\sum_n\chi_\beta\al{p,n}\,\ee^{\zeta_\beta\al{p,n}\, t}\ ,
\end{eqnarray}
with
$\chi_\alpha\al{p,n}$, $\chi_\beta\al{p,n}$, $\zeta_\alpha\al{p,n}$ and $\zeta_\beta\al{p,n}$ time-independent complex coefficients,
whose specific form can be computed iteratively.
Moreover, the expression for $\alpha\al{p}(t)$ and $\beta_j\al{p}(t)$ are found integrating Eq.~\rp{dotalphabeta} and are given by
\begin{eqnarray}\label{solalphabeta}
\alpha\al{p}(t)&=&\sum_n\frac{\chi_\alpha\al{p,n}}{z+\zeta_\alpha\al{p,n}} \
\pt{\ee^{\zeta_\alpha\al{p,n}\, t}-\ee^{-z\,t}},
\nn\\
\beta_j\al{p}(t)&=&\sum_n \frac{\chi_\beta\al{p,n}}{w_j+\zeta_\beta\al{p,n}}\
\pt{ \ee^{\zeta_\beta\al{p,n}\, t}-\ee^{-w_j\,t}}\ .
\end{eqnarray}
We note that all the coefficients $\zeta_\alpha\al{p,n}$ and $\zeta_\beta\al{p,n}$ have non-positive real parts, ${\rm Re}\pq{\zeta_\alpha\al{p,n}}, {\rm Re}\pq{\zeta_\beta\al{p,n}}\leq 0$, thus the large-time solutions $\alpha\al{p}(t\to\infty)\equiv\alpha_{st}\al{p}(t)$ and $\beta_{j}\al{p}(t\to\infty)\equiv\beta_{j,st}\al{p}(t)$ are found from Eq.~\rp{solalphabeta} by keeping only the terms
for which $\zeta_\alpha\al{p,n}$ and $\zeta_\beta\al{p,n}$ are purely imaginary, that can be shown to be equal to $\ii\,n\,\omega_+$ with $n$ odd and even integer  respectively. In particular
$\alpha_{st}\al{p}(t)$ and $\tilde\beta_{j,st}\al{p}(t)$ are periodic functions (with period $2\pi/\omega_+$) which contains frequency components that are, respectively, odd and even multiples of $\omega_+$,
\begin{eqnarray}\label{solalphabeta0}
\alpha_{st}\al{p}(t)&=&\sum_\pp{n=-p-1}{n\ {\rm odd}}^{p+1}\frac{\tilde\chi_\alpha\al{p,n}}{z+i\,n\,\omega_+} \
\ee^{i\,n\,\omega_+\, t},
\nn\\
\beta_{j,st}\al{p}(t)&=&\sum_\pp{n=-p-1}{n\ {\rm even}}^{p+1} \frac{\tilde\chi_\beta\al{p,n}}{w_j+i\,n\,\omega_+}\
 \ee^{i\,n\,\omega_+\, t},
\end{eqnarray}
where $z$ and $w_j$ are defined in Eq.~\rp{zw}, and $\tilde \chi_\alpha\al{p,n}(t)$ and  $\tilde\chi_\beta\al{p,n}(t)$ are the coefficients that correspond to those particular parameters $\zeta_\alpha\al{p,n}$ and $\zeta_\beta\al{p,n}$ that are imaginary.

\subsection{Resonant and non-resonant terms}

The QLE, in the interaction picture with respect to the Hamiltonian
$\hat{H}_0=\hbar\pt{\omega_-a\da a+\omega_1 \hat{b}^\dagger_1 \hat{b}_1+\omega_2 \hat{b}^\dagger_2 \hat{b}_2}$,
reduce to
\begin{eqnarray}\label{qle1_b0}
\dot{\delta\hat{a}}&=&-\pt{\kappa + i \Delta_0} \delta \hat{a}+\sqrt{2\kappa} \hat{a}^{\rm in}
\\&&
+2i \pt{g_1{\rm Re}\pq{\beta_1(t)}+ g_2{\rm Re}\pq{\beta_2(t)}} \delta \hat{a}
  \nn \\&&
  -i\alpha(t)\ee^{i\omega_- t}
\lpq{g_1\pt{\delta\hat b_1\ee^{-i\omega_1t}+\delta\hat b_1\da\ee^{i\omega_1t}}
}\nn\\&& \rpq{
+g_2\pt{\delta\hat b_2\ee^{-i\omega_2t}+\delta\hat b_2\da\ee^{i\omega_2t}}},
\nn\\
\dot{\delta \hat{b}}_j&=&-{\frac{\gamma_j}{2}} \delta \hat{b}_j+\sqrt{\gamma_j} \hat{b}_j^{\rm in}
-ig_j\ee^{i\omega_jt}\pq{
\ee^{i\omega_- t}\alpha(t)
\delta\hat a\da+H.c.}\ .
\nn
\label{qle2_b0}
\end{eqnarray}
Before proceeding, we note that we can include the DC component of $\beta_j(t)$ into the cavity detuning, hence we introduce
$\Delta=\Delta_0+2\sum_{j=1,2} g_j{\rm Re}\pq{\sum_p g^p\frac{\tilde\chi_\beta\al{p,0}}{w_j}}$, according to the notation introduced in Eq.~\rp{barDelta},
\begin{eqnarray}
\sum_p g^p \frac{\tilde\chi_\beta\al{p,0}}{w_j}\equiv \beta_j^{\rm DC}\ .
\end{eqnarray}
 Moreover we can isolate the resonant terms of the QLE, namely the terms with time-independent coefficients, by considering the  lowest order frequency components of $\alpha(t)$, i.e.,
 \begin{eqnarray}\label{alpha_pm}
\alpha_\pm=\sum_p g^p\frac{\tilde\chi_\alpha\al{p,\pm 1}}{z\pm i\,\omega_+},
\end{eqnarray}
corresponding to the frequencies $\pm \omega_+$,
and defining
\begin{eqnarray}\label{barbeta}
\bar\beta_j(t)&=&\beta_j(t)-\beta_j^{\rm DC},
\\
\bar\alpha_+(t)&=&\alpha(t)-\ee^{i\omega_+t}\alpha_+,
\nn\\
\bar\alpha_-(t)&=&\alpha(t)-\ee^{-i\omega_+t}\alpha_-\ .
\label{baralpha}
\end{eqnarray}
Thereby we find
\begin{eqnarray}\label{qle1_b1}
\dot{\delta\hat{a}}&=&-\pt{\kappa + i \Delta} \delta \hat{a}
 -i\alpha_- g_1\delta\hat b_1\da
 -i\alpha_+  g_2 \delta\hat b_2
 \nn\\&&
+\sqrt{2\kappa} \hat{a}^{\rm in}+F_a(t),
 \nn\\
\dot{\delta \hat{b}}_1&=&-{\frac{\gamma_1}{2}} \delta \hat{b}_1
-ig_1\alpha_-\delta\hat a\da
+\sqrt{\gamma_1}  \hat{b}_1^{\rm in} +F_{b_1}(t),
 \nn\\
\dot{\delta \hat{b}}_2&=&-{\frac{\gamma_2}{2}} \delta \hat{b}_2
-ig_2{\alpha_+}^*\delta\hat a
+\sqrt{\gamma_2} \hat{b}_2^{\rm in} +F_{b_2}(t),
\label{qle2_b1}
\end{eqnarray}
where $F_a(t), F_{b_1}(t)$ and $F_{b_2}(t)$ account for the terms with time-dependent coefficients and are given by
\begin{eqnarray}\label{FF}
F_a(t)&=&+2i \pt{g_1{\rm Re}\pq{\bar\beta_1(t)}+ g_2{\rm Re}\pq{\bar\beta_2(t)}} \delta \hat{a}
  \nn \\&&
 -i\alpha(t)\ee^{i\omega_- t}
\pt{g_1\delta\hat b_1\ee^{-i\omega_1t}
+g_2\delta\hat b_2\da\ee^{i\omega_2t}}
 \nn\\&&
 -i\bar\alpha_-(t)\ee^{i\omega_- t} g_1\delta\hat b_1\da\ee^{i\omega_1t}
 -i\bar\alpha_+(t)\ee^{i\omega_- t}g_2 \delta\hat b_2\ee^{-i\omega_2t},
\nn\\
F_{b_1}(t)&=&-ig_1\ee^{i\omega_1t}\pq{
\ee^{i\omega_- t}\bar\alpha_-(t)
\delta\hat a\da+\ee^{-i\omega_- t}\alpha^*(t)
\delta\hat a},
\nn\\
F_{b_2}(t)&=&-ig_2\ee^{i\omega_2t}\pq{
\ee^{i\omega_- t}\alpha(t)
\delta\hat a\da+{\ee^{-i\omega_- t}\bar\alpha_+(t)}^*
\delta\hat a}\ .
\end{eqnarray}
In particular we can introduce the linearized coupling strength $G\al{tot}_1=g\alpha_-$ and $G\al{tot}_2=g\alpha_+$, with $\alpha_\pm$ defined in Eq.~\rp{alpha_pm}. The expressions introduced in Eq.~\rp{G1G2} correspond to the expansion of these parameters at zeroth order in $g$ (see also Eq.~\rp{firstorder}).

We are interested in the regime in which the terms in Eqs.~\rp{FF} with time-dependent coefficients are negligible. They can be neglected when
$g_j\,\abs{\alpha_{st}(t)}, g_j\,\abs{\bar\beta_{j,st}(t)}, \,\kappa \ll{\rm min}\pg{\omega_1,\omega_2,\abs{\omega_1-\omega_2}}.$
In particular this condition is true when it is valid for the lowest order term in the expansion in power of $g$. In details, the non resonant terms can be neglected when
\begin{eqnarray}\label{cond1}
g\, \abs{\alpha_{\pm}},\, \kappa \ll{\rm min}\pg{\omega_1,\omega_2,\abs{\omega_1-\omega_2}}\ .
\end{eqnarray}
When this condition is fulfilled the parameters  $\alpha(t)$ and $\beta_j(t)$ can be safely expanded at the lowest order in $g$. Specifically they can be approximated as
 \begin{eqnarray}\label{firstorder}
 \alpha_-&\simeq&\frac{-i E_1}{z-i\omega_+}\ ,
\ \ \ \ \ \ \ \ \  \ \alpha_+\simeq\frac{-i E_2}{z+i\omega_+}\ ,
\\
\alpha_{st}(t)&\simeq&\alpha_{st}\al{0}(t)= \alpha_-\ee^{-i\omega_+t}+\alpha_+\ee^{i\omega_+t},
\nn\\
\beta_j^{\rm DC}&\simeq& \frac{-i g_j}{w_j}\pq{ \alpha_-{\alpha_-^*}+ \alpha_+{\alpha_+^*}},
\nn\\
\bar\beta_{j,st}(t)&\simeq&g_j\,\beta_{j,st}\al{1}(t)-\beta_j^{\rm DC}
\nn\\&=&-i g_j\pq{
\frac{ \alpha_-{\alpha_+^*}}{w_j-2i\omega_+}\ee^{-2i\,\omega_+\,t}
\frac{ \alpha_+{\alpha_-^*}}{w_j+2i\omega_+}\ee^{2i\,\omega_+\,t}
}\ . \nn
\end{eqnarray}
Moreover the parameters $\bar\alpha_\pm(t)$ defined in Eq.~\rp{baralpha} are zero.
Using these expressions the QLE in Eq.~\rp{qle1_b1} can be rewritten as
\begin{eqnarray}\label{qle1_b2}
\dot{\delta\hat{a}}&=&-\pt{\kappa + i \Delta} \delta \hat{a}
 -iG_1\delta\hat b_1\da
 -iG_2 \delta\hat b_2
 \nn\\&&
+\sqrt{2\kappa} \hat{a}^{\rm in}+F_a(t),
\nn\\
\dot{\delta \hat{b}}_1&=&-{\frac{\gamma_1}{2}} \delta \hat{b}_1
-iG_1\delta\hat a\da
+\sqrt{\gamma_1} \hat{b}_1^{\rm in} +F_{b_1}(t),
 \nn\\
\dot{\delta \hat{b}}_2&=&-{\frac{\gamma_2}{2}} \delta \hat{b}_2
-iG_2^*\delta\hat a
+\sqrt{\gamma_2} \hat{b}_2^{\rm in} +F_{b_2}(t),
\label{qle2_b2}
\end{eqnarray}
with $G_1$ and $G_2$ defined in Eq.~\rp{G1G2} and
\begin{eqnarray}
F_a(t)&\simeq&+2i \pt{g_1^2{\rm Re}\pq{\bar\beta_{1,st}\al{1}(t)}+ g_2^2{\rm Re}\pq{\bar\beta_{2,st}\al{1}(t)}} \delta \hat{a}
  \nn \\&&
 -i\ee^{i\,\omega_-\,t}\alpha_{st}\al{0}(t)
\pt{g_1\delta\hat b_1\ee^{-i\omega_1t}
+g_2\delta\hat b_2\da\ee^{i\omega_2t}},
\nn\\
F_{b_1}(t)&\simeq&-ig_1\ee^{i(\omega_1+\omega_-)t}{\alpha_{st}\al{0}}^*(t)\delta\hat a,
\nn\\
F_{b_2}(t)&\simeq&-ig_2\ee^{i(\omega_2+\omega_-)t}
\alpha_{st}\al{0}(t)
\delta\hat a\da\ .
\end{eqnarray}
When the time-dependent coefficients are neglected these equations reduce to Eqs.\rp{deltaa2}--\rp{b2st}.

\end{document}